\documentstyle[12pt,epsfig,aps]{revtex}

\begin{document}

\draft

\title{Non--perturbative quenched propagator beyond the infrared approximation}

\author{J. S\'anchez--Guill\'en, R.A.~V\'azquez}

\address{Departamento de F\'{\i}sica de Part\'{\i}culas, \\
Universidad de Santiago, \\ 15706 Santiago de Compostela, Spain.}

\date{\today}
\maketitle

\begin{abstract}
A new approach to the quenched propagator in QED beyond the IR limit is 
proposed. The method is based on evolution equations in the proper time.
\end{abstract}

\section{Introduction}

We propose a new non-perturbative analysis of quenched QED beyond the
infrared Bloch Nordsieck approximation using the world line formalism.
The main idea is to formulate evolution equations in the proper time, inspired
by the rather intuitive resulting classical picture. As a warm up we obtain
in a simple way  the renormalized quenched propagator in the infrared (IR)
limit.
From there we procced to improve on this IR approximation, which we achieve in
the form of integro--differential equations, our main results.  They
are of course difficult to solve, but they have an appealing physical
interpretation and suggest a simplified numerical analysis.

The paper is organized as follows. In section \ref{infrared} we study the
infrared behavior of lepton propagators for several cases. First we
consider the Bloch Nordsieck approximation as given by uniform proper
time paths, which exhibit the classical image in the worldline.
This suggests the use of the proper time evolution equations as a short
cut to go beyond the infrared approximation. The idea is carried out for
scalar QED and then for the full Dirac algebra.
The lesson of this section is a better understanding of
renormalization and the fact that
the scalar case contains the basic physics, which will be used
in the next section, section \ref{beyond} where, we develop further the
proper time formalism to go beyond the infrared limit. The results are
summarized in section \ref{conclu} devoted to the conclusions. 

\section{Infrared Behavior: the classical image of the worldline formalism}
\label{infrared}

In this section we will study the classical Block--Nordsieck approximation and
their extensions for scalar and fermionic QED. Our main point is given 
in Equation (\ref{evol1}), where an evolution equation is introduced
for the quenched propagator. Its 
solution is the well--known Block--Nordsieck result. The extensions to the 
scalar and full Dirac algebra are carried out without difficulty. 

\subsection{Bloch--Nordsieck approximation}
\label{bn}

The Bloch--Nordsieck model has been repeatedly used to study the
infrared behavior of QED. Here we will consider it from the path
integral point of view as a convenient starting point.
The model has been extensively studied in the literature
\cite{BN,Weinberg_1,Harada,Kibble,Karanikas_1}

In the Bloch--Nordsieck model one substitutes the Dirac's gamma
matrices by a constant vector $\gamma_\mu \rightarrow u_\mu$. This
constant vector will be identified to the (constant) velocity of
the electron, see below. With this substitution the model is
completely integrable and the electron Green function can be cast
in terms of elementary functions.
The electron propagator in the Bloch--Nordsieck approximation is
given by
\begin{equation}
[u_\mu (i \frac{\partial}{\partial x_\mu}+e A_\mu(x) ) - m ]
G(x,y) = -\delta(x-y),
\end{equation}
where $A_\mu(x)$ is the electromagnetic potential. We will solve this
model by introducing a proper time (Schwinger's method
\cite{Schwinger}) \cite{Barbashov,Fradkin,Karanikas_1,Karanikas_2}.
Define
\begin{equation}
G(x,y)=i \int_0^\infty d\tau G(\tau,x,y),
\end{equation}
then $G(\tau,x,y)$ verifies the equation
\begin{equation}
-i \frac{\partial G}{\partial \tau} =
[u_\mu (i \frac{\partial}{\partial x_\mu}+e A_\mu(x) ) - m ]
G(\tau,x,y) = {\cal H} G(\tau,x,y),
\label{bn_equ}
\end{equation}
with initial conditions
\begin{equation}
G(\tau=0,x,y) = \delta(x-y).
\end{equation}
The formal solution is given by
\begin{equation}
G(\tau,x,y) = \exp\{-i \tau {\cal H}\} \;\; \delta(x-y).
\end{equation}
Here ${\cal H}$ is clearly seen as a Hamiltonian which gives the
evolution in the proper time of the Schr\"odinger--like equation.
This formal expression can be cast into a path integral in the
proper time. We can write
\begin{eqnarray}
G(\tau,x,y)& = & N \int Dx(\tau) Dp(\tau) \exp\{ i \int_0^\tau d\tau
(p \dot x - {\cal H}(x,p))\} \nonumber \\
  & = & N \int Dx(\tau) Dp(\tau) \exp\{ i \int_0^\tau d\tau
(p \dot x - u p - e u A(x) +m)\}.
\end{eqnarray}
The electromagnetic field is coupled to the current
$j_\mu(z) = u_\mu \int_0^\tau d\tau \delta(z-x(\tau))$ and can be
integrated out, since the resulting integral is gaussian
\begin{equation}
<G(\tau,x,y)> = \int DA_\mu \; G(\tau,x,y) \; \exp\{-i
\int dx \frac{1}{4} F^2 +\frac{\lambda}{2} (\partial A)^2\},
\end{equation}
where $\lambda$ is a gauge--fixing parameter and the brackets $<>$
mean the average over the gauge fields.

The integration of the $A$ fields gives a non-local lagrangian for
$<G>$
\begin{eqnarray}
<G(\tau,x,y)> & = & N \int Dx Dp \exp\{i \int_0^\tau d\tau (p \dot x - p u +m)
\}
\nonumber \\
& &
\exp\{-\frac{e^2}{2} \int_0^\tau d\tau_1 \int_0^\tau d\tau_2
\int \frac{dk}{(2\pi)^D} u_\mu u_\nu D_{\mu \nu}(k)
e^{-i k(x(\tau_1)-x(\tau_2))}\},
\end{eqnarray}
$D_{\mu \nu}(k)$ is the photon propagator in the gauge $\lambda$.

The integral over $p$ gives a delta function,
$\delta(\dot x_\mu(\tau)-u_\mu))$, which implies that the particle
moves with constant velocity, $u_\mu$, as stated previously.
We see that the electron in the Bloch--Nordsieck approximation behaves as a
classical particle, the only path that contributes to the propagator
is the classical path and quantum fluctuations are {\it exactly }
canceled. This fact is most easily seen in the above formalism.
The triviality of the quantum corrections is due to the appearance
of the delta function, which in turn is related to the triviality
of the classical hamiltonian. For the BN model,
the classical path and the uniform velocity path coincide,
independently of the external potential, $A_\mu(x)$.
Alternatively one may notice that the Eq.(\ref{bn_equ}) is first
order on the variable $x$. As is well known from the theory of
partial differential equations, the solution admits a particle
interpretation. This is not true for the cases considered below. The uniform 
path is only an approximation valid when the momentum interchanged with the 
gauge fields is sufficiently small such that the path can be considered 
uniform, {\it i.e.} in the infrared domain.

Therefore the non-local term must be evaluated in the classical path
alone. It gives
\begin{equation}
{\rm N.L.} = \frac{3-\lambda}{2} \int_0^\tau d\tau_1 d\tau_2
\frac{1}{(\tau_1-\tau_2)^2}.
\end{equation}
After renormalization we find
\begin{equation}
<G(\tau,p)> = i \tau^a \exp\{-i \tau (m - up)\},
\label{bn_end}
\end{equation}
where $a=(3-\lambda)/(2 \pi) \alpha$. Integrating over $\tau$, we
obtain finally
\begin{equation}
<G(p)> = i \int_0^\infty d\tau <G(\tau,p)> = \Gamma(1+a)
\frac{i}{(m-up)^{1+a}}.
\end{equation}
This is the well--known Bloch--Nordsieck result. In the Yennie-Suura
gauge, $\lambda=3$, the electron propagator reduces to the free case.
Notice that $<G(p,\tau)>$ verifies the homogeneous evolution equation
\begin{equation}
-i \frac{\partial <G(p,\tau)>}{\partial \tau} = (up -m ) <G(p,\tau)> 
+ \frac{a}{\tau} <G(p,\tau)>,
\label{evol1} 
\end{equation}
which has of course the Eq.(\ref{bn_end}) above as a solution. 
This fact will be used later on to go beyond the infrared approximation.

\subsection{Scalar QED}

Before considering the spinor electrodynamics, let's study the
somewhat simpler case of scalar QED. We will study in detail the
renormalization for this case.

The scalar propagator, coupled to the electromagnetic current, is
given by
\begin{equation}
((\partial_\mu - e A_\mu(x) )^2 -m^2 ) G(x,y) = \delta(x-y),
\end{equation}
as before we introduce the Schwinger proper time
\begin{equation}
i \frac{\partial G(\tau,x,y)}{\partial \tau} =
((\partial_\mu - e A_\mu(x) )^2 -m^2 ) G(x,y),
\end{equation}
$G(\tau,x,y)$ has the solution
\begin{equation}
G(\tau,x,y)= \exp\{-i {\cal H} \tau \} \, \delta(x-y)
\label{scalar_prop}
\end{equation}
which can be expressed as a path integral
\begin{eqnarray}
G(\tau,x,y) & = & N \int Dx(\tau) Dp(\tau) \exp\{ i \int_0^\tau d\tau
(p \dot x + {\cal H}(x,p))\} \nonumber \\
&  = &  N \int Dx(\tau) Dp(\tau) \exp\{ i \int_0^\tau d\tau
(p \dot x + (p + e A(x))^2 -m^2)\}.
\end{eqnarray}
Now $p$ appears quadratically and the integral over it is no longer
a delta function. As expected, the classical path is not the only contribution
to the path integral and quantum corrections are relevant.
However, for the moment, we will consider the contribution given by the uniform
path alone, as the main contribution in the infrared limit.
We first shift the momentum integral $\Pi = p+ e A(x)$
\begin{equation}
G(\tau,x,y)=
N \int Dx(\tau) D\Pi(\tau) \exp\{ i \int_0^\tau d\tau
(\Pi \dot x + \Pi^2 -m^2 - e \dot x A_\mu(x))\}.
\end{equation}
and as before, the electromagnetic field is coupled to the
current $j_\mu(z) = \int d\tau \dot x_\mu(\tau) \delta(x(\tau)-z)$.
The average over gauge fields gives therefore, the same non local
term
\begin{equation}
{\rm N.L.} = \int d\tau d\tau' \int \frac{dk}{(2\pi)^D} \dot x_\mu(\tau)
\dot x_\nu(\tau')\, \frac{ D_{\mu\nu}(k)}{k^2} \, e^{i k(x(\tau)-x(\tau'))}.
\end{equation}
Now, we make the uniform path approximation, $\dot x_\mu(\tau) =
u_\mu =$ const. which gives
\begin{equation}
{\rm N.L.} = \int d\tau d\tau' \int \frac{dk}{(2\pi)^D} u_\mu
u_\nu\frac{ D_{\mu\nu}(k)}{k^2} e^{i ku(\tau-\tau')}.
\end{equation}
Integrating over $\tau$ and $\tau'$ gives
\begin{equation}
{\rm N.L.} = - \int \frac{dk}{(2\pi)^D} u_\mu
u_\nu\frac{ D_{\mu\nu}(k)}{k^2} \frac{1-\cos(ku \tau)}{(k u)^2}.
\end{equation}
The integral is IR finite, at low momentum the divergent denominator 
cancels out with the $1-\cos(ku \tau)$ in the numerator. 
We will renormalize the ultraviolet divergent $k$ integral using
dimensional regularization, expanding around four dimensions,
$D=4-\epsilon$, we obtain \cite{Karanikas_1}
\begin{equation}
{\rm N.L.} = -\frac{A}{\epsilon} (\mu^2 \tau)^{\epsilon},
\end{equation}
where
\begin{equation}
A= \frac{1}{8 \pi^{D/2} } \Gamma(\frac{D}{2}-1) \frac{(1-(-1)^{2-D})}{3-D} (1-
\frac{\lambda}{2} (3-D)) = A_0 + \epsilon A_1 + \ldots,
\end{equation}
and $A_0=-1/(4 \pi^2)$.
The scalar propagator can be written
\begin{equation}
<G(x,y)> = \int \frac{dp}{(2 \pi)^D} e^{i p(x-y)} \int_0^\infty d\tau
\exp{-i \tau (m^2-p^2)} \, \exp{-e^2 \frac{A}{\epsilon} (\mu^2 \tau)^\epsilon}.
\end{equation}
Expanding the exponential in powers of $e$ we obtain
\begin{equation}
<G(p,\tau)> = \int_0^\infty d\tau \exp{-i \tau (m^2-p^2)} \sum_{n=0}^\infty
(\frac{-e^2 A}{\epsilon})^n (\mu^2 \tau)^{n \epsilon}.
\end{equation}
Each term in the series can be integrated and gives
\begin{equation}
<G(p,\tau)> = \frac{-i}{(m^2-p^2)} \sum_{n=0}^\infty \frac{(-e^2 A)^n}{n!
\epsilon^n} \Gamma(1+n \epsilon) (\frac{2 i \mu^2}{m^2-p^2})^{n \epsilon}
\end{equation}
The $n$--th term can be written as the $n$-th power of
\begin{equation}
-e^2 \frac{A}{\epsilon} e^{\epsilon \log((2 i \mu^2)/(m^2-p^2))}
  \Gamma(1+n \epsilon)^{1/n},
\end{equation}
Expanding now in powers of $\epsilon$ we arrive at
\begin{equation}
-e^2 \left( \frac{A_0}{\epsilon} + A_1 - A_0 \gamma + A_0 \log((2 i
  \mu^2)/(m^2-p^2)) + {\cal O}(\epsilon) \right),
\end{equation}
and we can choose the counterterms in such a way to cancel
the term $A_0/\epsilon + A_1 - A_0 \gamma$. The scalar propagator
after renormalization is given by
\begin{equation}
<G_R(p)> = \frac{i}{m^2-p^2} e^{-a \log(\mu^2/(m^2-p^2))},
\end{equation}
which coincides with the Block-Nordsieck result, as expected.
Notice that care has to be taken in the order of the limit $\epsilon
\rightarrow 0$ and
the integral over $\tau$. We will see next that 
including the Dirac algebra is under control, this will allow us to go back
to the scalar case for the purposes of the present article. 

\subsection{Full Dirac algebra}

Including the full Dirac algebra is far from trivial. It has been
done in references \cite{Fradkin,Alexandrou,Schmidt,Karanikas_2}.

We will follow the method of Ref.\cite{Alexandrou}. 
The Dirac equation is
\begin{equation}
(i \gamma_\mu \partial_\mu - e \gamma_\mu A_\mu(x) -m ) G(x,y) = \delta(x-y).
\end{equation}
Introducing the proper time as before, (see Eq.(6) from ref.\cite{Alexandrou})
\begin{equation}
G(x,y) = \int_0^\infty dT \int d\chi \exp\{-\frac{i}{2} (m^2 T + m \chi)\}
\exp\{\frac{i}{2} ( (\gamma \Pi)^2 T + \gamma \Pi \chi)\},
\end{equation}
where $\Pi^\mu = p^\mu -e A^\mu(x)$ and $\chi$ is a Grassman variable.
Introducing a set of auxiliary Grassman variables $\xi_\mu$ we can write a
path integral for the propagator
\begin{equation}
G(x,y)= \exp\{\gamma \frac{\partial}{\partial \Gamma} \}
\int_0^\infty dT \int d\chi \int Dx Dp D\xi \exp\{i \int_0^T
d\tau (p \dot x + i \xi \dot \xi - {\cal H}) - \xi(0) \xi(T)\},
\end{equation}
where ${\cal H}={1 \over 2} (-\Pi^2 + 2 i e F_{\mu \nu} \xi^\mu \xi^\nu - 
{2 \over T} \, \Pi_\mu \xi_\mu \chi)$. The $\xi_\mu$ verifies the boundary 
conditions $\xi_\mu(0) + \xi_\mu(T) = \Gamma_\mu$. Shifting the momentum as 
before and integrating
the resulting gaussian integral over $\Pi$ we obtain
\begin{equation}
G(x,y)= \exp\{\gamma \frac{\partial}{\partial \Gamma} \}
\int_0^\infty dT \int d\chi {\cal N}(T) \exp\{-\frac{i}{2} (m^2 T +m \chi)\}
\int Dx D\xi \exp\{i S[x,\xi]\},
\end{equation}
where $S[x,\xi]$ is given by
\begin{equation}
S[x,\xi]= \int_0^T d\tau ( -\frac{1}{2} \dot x^2 + i \xi \dot \xi -
i e F_{\mu \nu} \xi^\mu \xi^\nu -\frac{1}{T} \dot x \xi \chi)
-i \xi(0) \xi(T).
\end{equation}
As before the gauge fields appear linearly only and the integration over
them can be done exactly.
\begin{equation}
<G(x,y)>= \exp\{\gamma \frac{\partial}{\partial \Gamma}\}
\int_0^\infty dT \int d\chi {\cal N}(T) \int Dx D\xi \exp\{i S_{\small
\rm eff}\},
\end{equation}
where the action $S_{\rm eff}$ is given by
\begin{eqnarray}
S_{\small \rm eff} & = & \int_0^T d\tau (-\frac{1}{2} (\dot x)^2 + i
\xi \dot \xi ) \nonumber \\
& & \frac{-e^2}{2} \int_0^T d\tau_1 d\tau_2 \int \frac{dk}{(2 \pi)^D}
\frac{D_{\mu \nu}(k)}{k^2}
\left( \dot x^1_\mu + 2 \xi^1_\mu k.\xi^1 \right) \left( \dot x^2_\mu - 2
\xi^2_\mu k.\xi^2 \right) e^{-i k(x^1-x^2)},
\end{eqnarray}
$x^1 = x(\tau_1)$ and so on.

We will calculate the action for the uniform path, as done before.

{From} the classical equations of movement for the free case ($e=0$) we get
that the uniform path in our case is given by
\begin{eqnarray}
x_\mu(\tau)=  & x_\mu(0) + p_\mu \tau + \frac{\tau}{T} \xi_\mu(0) \chi,
\nonumber \\
\xi_\mu(\tau) = & \xi_\mu(0) - \frac{\tau}{2 i T} p_\mu \chi.
\end{eqnarray}
Introducing this solution in the path integral and after some algebra
we arrive at
\begin{eqnarray}
<G(x,y)> & = & \exp\{\gamma \frac{\partial}{\partial \Gamma} \}
\int \frac{dp}{(2 \pi)^4} e^{i p (x-y)} \nonumber
\int_0^\infty dT \int d\chi \exp\{\frac{-i}{2} (m^2 T + m \chi)\} \\
& &
\exp\{\frac{i}{2} p^2 T + i p
\xi(0) \chi - e^2 \frac{A}{\epsilon} (\mu T p )^{\epsilon} ( 1 +
\frac{p \xi(0) \chi}{p^2 T} \epsilon)\},
\end{eqnarray}
where $A$ is given as before
Integrating over $\chi$ and applying the $\partial/\partial\Gamma$
we arrive at
\begin{eqnarray}
<G(x,y)> & =& \int \frac{dp}{(2 \pi)^4} e^{i p (x-y)}
\int_0^\infty dT \exp\{\frac{-i}{2} (m^2 T)\} \exp\{\frac{i}{2} p^2 T -
e^2 \frac{A}{\epsilon} (\mu T p )^{\epsilon} \} \nonumber \\
& & \left( \frac{-i}{2} (m + p\gamma) + \frac{e^2 A}{2 \epsilon} (\mu T
p)^{\epsilon} \epsilon \frac{p \gamma}{p^2 T} \right).
\end{eqnarray}
If we set $e=0$ we obviously reproduce the free fermion
propagator. For $e\ne0$ care must be taken in the regularization
procedure since the integration over $T$ and the limit $\epsilon
\rightarrow 0$ do not commute. If we boldly take the limit first and
integrate we arrive at
\begin{equation}
<G(p)> = Z \frac{(\mu p)^a}{(p^2-m^2)^a} \Gamma(1+a) \frac{1}{p^2-m^2}
\left( m + p\gamma \left( 1 + \frac{p^2-m^2}{2 p^2} \right) \right),
\end{equation}
where $Z$ is the renormalization constant $Z=\exp(-e^2/\epsilon A +\ldots)$

Taking correctly the limit after the integration we arrive at
\begin{equation}
<G(p)> = Z \frac{(\mu p)^a}{(p^2-m^2)^a} \Gamma(1+a) \frac{1}{p^2-m^2}
\left( m + p\gamma \right),
\end{equation}
which is again the Block--Nordsieck result and reproduces the IR logs upon
expansion on powers of $a$. 

\section{Beyond Infrared Behavior: proper time evolution equations}
\label{beyond}

We will now write down the evolution equations for a scalar coupled to 
a scalar field. We will use this simpler case to demonstrate that the 
evolution equations allow us to go beyond the infrared domain. The obvious
next step would be to construct evolution equations whose perturbative 
expansion give exact results up to a given order. Let's start by writing
down equations valid up to first order. Further
generalizations to the spinor case and higher orders 
will be left for future work.

As before we start with the scalar propagator as given in 
Eq.(\ref{scalar_prop}) in the proper time formalism, 
but now for a scalar field coupled to a scalar field 
${\cal H}= \partial_\mu^2 - m^2 + g\phi(x)$. The propagator
verifies the evolution equation
\begin{equation}
-i \frac{\partial G}{\partial \tau} = [ (\partial_\mu^2 -m^2) + g \phi(x) ]
G(\tau,x,y),
\end{equation}
where $g$ is the coupling constant to the scalar field $\phi(x)$. We want to 
average over the field $\phi$ with a weight given by the free action.
\begin{equation}
<{\cal O}> = \int d\phi {\cal O} \exp\{i \int dx {\cal L_\phi} \},
\end{equation}
${\cal L_\phi}= \phi (\partial_\mu^2 -m^2_\phi) \phi$. Therefore we get
\begin{equation}
-i \frac{\partial <G>}{\partial \tau} =  (\partial_\mu^2 -m^2) <G> + 
g < \phi(x) G(\tau,x,y) >.
\end{equation}
In the infrared limit the average
$< \phi(x) G(\tau,x,y) >$ is simply given by $a/\tau \,G$, as shown
above by Eq.(\ref{evol1}). But in general it can obviously not
be written as a simple operator acting on $G$. So one has to estimate the 
value of 
$< \phi(x) G(\tau,x,y) >$ by introducing as before a path integral 
representation for $G$.
\begin{equation}
G(\tau,x,y) = N \int Dx Dp \exp\{ -i \int_0^\tau d\tau p \dot x + p^2 -m^2 +
g \phi(x) \},
\end{equation}
then
\begin{eqnarray}
<\phi(x) G(\tau,x,y)> & = & \int D\phi Dx Dp \phi(x) 
\exp\{ -i \int_0^\tau d\tau p \dot x + p^2 -m^2 +
g \phi(x) \}  \nonumber \\ 
& & \exp\{ i \int dx \phi(x) (p^2 -m^2_\phi) \phi(x) \},
\end{eqnarray}
introducing the current $j_0(z) = g \int d\tau \delta(z-x(\tau))$ the average 
can be written
\begin{eqnarray}
<\phi(x) G(\tau,x,y)> & = & \frac{\delta }{\delta j(x)} \int Dx Dp 
\exp\{ -i \int_0^\tau d\tau p \dot x + p^2 -m^2 \} \nonumber \\ 
& & \int D\phi \exp\{ i \int dx  \phi (p^2 -m^2_\phi) \phi+ j \phi \},
\end{eqnarray}
where the derivative is evaluated at $j=j_0$. The integral over $\phi$ is 
gaussian an after performing the functional derivative we arrive at
\begin{eqnarray}
<\phi(x) G(\tau,x,y)> & = & \int Dx Dp 
\exp\{ -i \int_0^\tau d\tau p \dot x + p^2 -m^2 \} \nonumber \\
& & \exp\{ \frac{i g^2}{2} \int d\tau_1 d\tau_3 D(x(\tau_1),x(\tau_2))\}
g \int_0^\tau d \tau_0 D(x,x(\tau_0)),
\end{eqnarray}
where $D(x,y)$ is the free scalar propagator of field $\phi$.
This expression can be simplified further if we make some approximations 
and constitutes the starting point to obtain evolution equations. 
To start with one can 
neglect the non--local term in the exponential since it is order $g^2$.
Then it is straightforward to evaluate the functional integral since 
all integrals are gaussian. After some algebra we arrive at
\begin{equation}
<\phi(x) G(\tau,x,y) > = g \int_0^\tau d \tau_0 \int dk_0 D(k_0) e^{i k_0 x}
\int Dx Dp \exp\{-i \int_0^\tau d\tau p^2 + p\dot x - m^2 +j x \},
\end{equation}
where $j(\tau)=-k_0 \delta(\tau-\tau_0)$. The integral over $x$ can be done 
and gives now a delta function $\delta(\dot p - j)$, i.e. the path is still
uniform but now presents a jump at $\tau= \tau_0$ where the momentum changes 
from $p$ to $p-k_0$. The electron emits a hard photon of momentum $k_0$. 
Evaluating the action at this specific path gives the total contribution, in 
this approximation
\begin{eqnarray}
<\phi(x) G(\tau,x,y) > & = & 
g \int_0^\tau d \tau_0 \int dk_0 D(k_0) e^{i (k_0-p) (x-y)} \nonumber \\
& &\exp\{ -i ( (p-k_0)^2 (\tau -\tau_0) - m^2 (\tau -\tau_0) +p^2 \tau_0 -m^2 
\tau_0 ) \}.
\end{eqnarray}
We have separated the contributions from $\tau-\tau_0$ and $\tau_0$ to make it
clear that they are both the free scalar propagators , i.e.
\begin{equation}
<\phi(x) G(\tau,x,y) > = g \int_0^\tau d \tau_0 \int dk_0 D(k_0) 
e^{i k_0 (x-y)} G_0(\tau_0,p) G_0(\tau-\tau_0,p-k_0).
\end{equation}
This result is valid up to first order on $g$. Then the evolution equation, 
valid up to this order is
\begin{equation}
-i \frac{\partial <G>}{\partial \tau} = (p^2 -m^2) <G> + g^2 
\int_0^\tau d \tau_0 \int dk_0 D(k_0) 
G_0(\tau_0,p) G_0(\tau-\tau_0,p-k_0).
\end{equation}
The solution of this equation up to order $g^2$ gives the exact one loop 
scalar propagator, as it should. A natural extension of this result would 
be to change the free propagators, $G_0$, in the integral by the exact ones $G$ with the hope 
that the main higher order contributions will be included in this way. 
We arrive, therefore, to our final evolution equation
\begin{equation}
-i \frac{\partial <G>}{\partial \tau} = (p^2 -m^2) <G> + g^2 
\int_0^\tau d \tau_0 \int dk_0 D(k_0) 
G(\tau_0,p) G(\tau-\tau_0,p-k_0).
\label{final_evol}
\end{equation}
The solution of this equation gives the exact result up to first order. We
think that higher orders are approximately taken into account. In the 
infrared limit, the equation reduces to the Block-Nordsieck case. 

\section{Conclusions}
\label{conclu}

\begin{figure}[hbt]
\centering
\mbox{\epsfig{figure=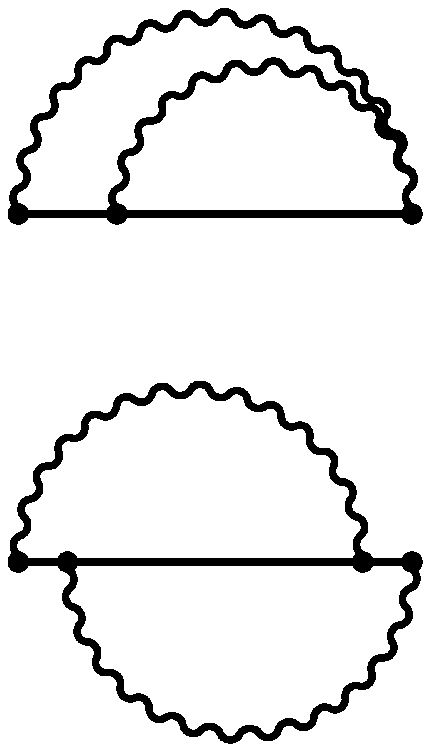,width=7.0cm}}
\caption{Diagrams included (upper graph) and not included (lower graph) on 
the evolution equation}
\label{fig1}
\end{figure}

We have arrived at an expression for the quenched propagator beyond the 
infrared
domain. It has a clear physical interpretation as one can see by looking
at the diagrams resummed by the above equation.
From direct inspection, one sees that the vertex function is evaluated 
exactly at the one loop level, and that there are an infinity number of such 
hard loops. Therefore, hard photons do not connect propagators with different
hard vertex, {\it i.e.} we have some kind of rainbow expansion, where 
overlapping loops are not included, as shown by Fig.\ref{fig1}.
But there are in addition an infinity number of infrared photons 
which join any two points in the propagator and make the infrared limit exact.
Since the proper time in the diagram is related to the virtuality of the 
propagator and we have, in the above equation, an integral from 0 to the upper
limit $\tau$, we expect that higher order loops are calculated only up to some
minimum virtuality, {\it i.e.} one expects an ordering in the virtuality 
of the diagrams that contribute to the above equation.

New insights in the nonperturbative regime are always welcome, even if 
quenched. In fact, better understanding of quenched approximations are
essential for lattice simulations. The non-positron approximation is also
useful in analytic attempts to fundamental problems, like stability of
relativistic QED, where renormalization is one of the main
problems \cite{Lieb}.
Our result in Eq.(\ref{final_evol}) can have  many practical 
applications like a systematic way to calculate hard loop corrections 
with exponentiation of soft photons. They will depend on our ability
to solve or approximate analytically the above equation. 
Of course Eq.(\ref{final_evol}) constitutes a new approach
and can serve in any case as a starting point for new numerical analysis.

{\bf Acknowledgements}
We thank Orlando Alvarez for discussions. One of us (RAV) thanks
Concha Gonzalez-Garcia for clarifying discussions. 
This work was supported by CICYT AEN-990589. 
R. V\'azquez is supported by the ``Ram\'on y Cajal'' fellowship program.

\end{document}